\documentstyle[11pt,newpasp,twoside,epsf]{article}
\def\edcomment#1{\iffalse\marginpar{\raggedright\sl#1\/}\else\relax\fi}
\reversemarginpar

\begin{document}
\title{Search For Exospheric Signatures From Transiting Planets}
 \author{N. Iro, A. Coustenis}
\affil{LESIA, Observatoire de Meudon,\\
 5 place J. Janssen, 92195 Meudon cedex, France}
\author{C. Moutou, N. Lajous}
\affil{Laboratoire d'Astrophysique de Marseille,\\
BP8, 13376 Marseille Cedex 12, France}
\author{M. Mayor, D. Queloz}
\affil{Observatoire de Gen\`{e}ve,\\
51 ch. des Maillettes, 1290 Sauverny, Switzerland}

\begin{abstract}
We searched for spectral signatures from the exosphere of HD209458 in the UV and the near-IR with UVES and ISAAC at the VLT.
We looked in  particular for the helium absorption feature at 10 830 \AA\, predicted to be among the strongest ones.
The upper limit of the He I line derived is 0.5\% at 3 $\sigma$ for a 3 \AA\ bandwith.
Planet-induced chromospheric activity search on HD209458 was also performed.
\end{abstract}

\section{Introduction}
The He I feature at 10 830 \AA\ is predicted to be one of the deepest absorption lines of the diffuse and extended exosphere in the optical and near-IR transmission spectrum The goal of the present study is to search for exospheric signatures in the spectrum of HD209458, during transits of the planet.
Shkolnik et al. (2003) suggested that a chomospheric activity can be induced by the planet orbit.
We searched for such a signature in stellar chromospheric lines.

\section{Search for spectral signatures from the exospheric components}
We performed observations of HD209458 in the UV with VLT/UVES in the 0.33-0.67 $\mu$m region at high spectral resolution.
1 transit was covered with R=70000 in December 1999 in two spectral range: Blue (0.328-0.4562 $\mu$m) and Red (0.4583-0.6686 $\mu$m).
6 transits were covered with R=100000:  1 in June 2000 in Blue and Red and June to September 2002 in Red (0.475-0.680 $\mu$m).
We searched for ions and neutral molecules originating in the planet's exosphere and located in the evaporated material around the planet.
Observations of 1 transit in the infrared was also performed with VLT/ISAAC at medium resolution (R=10000) around the He I line at1.083 $\mu$m (1.06-1.1 $\mu$m).
A sample of these spectra centered around this feature is shown is Fig. 1a.

From observations during one full transit with ISAAC in June 2001 of the spectral region around 1.083 $\mu$m,
no detection of He was obtained due to strong instrumental fringing affecting the data. 
An 3 $\sigma$ upper limit of 0.5\% deep absorption feature was derived.

\begin{figure}  
\begin{center}
\epsfclipon
\epsfysize=4.35cm
\epsffile{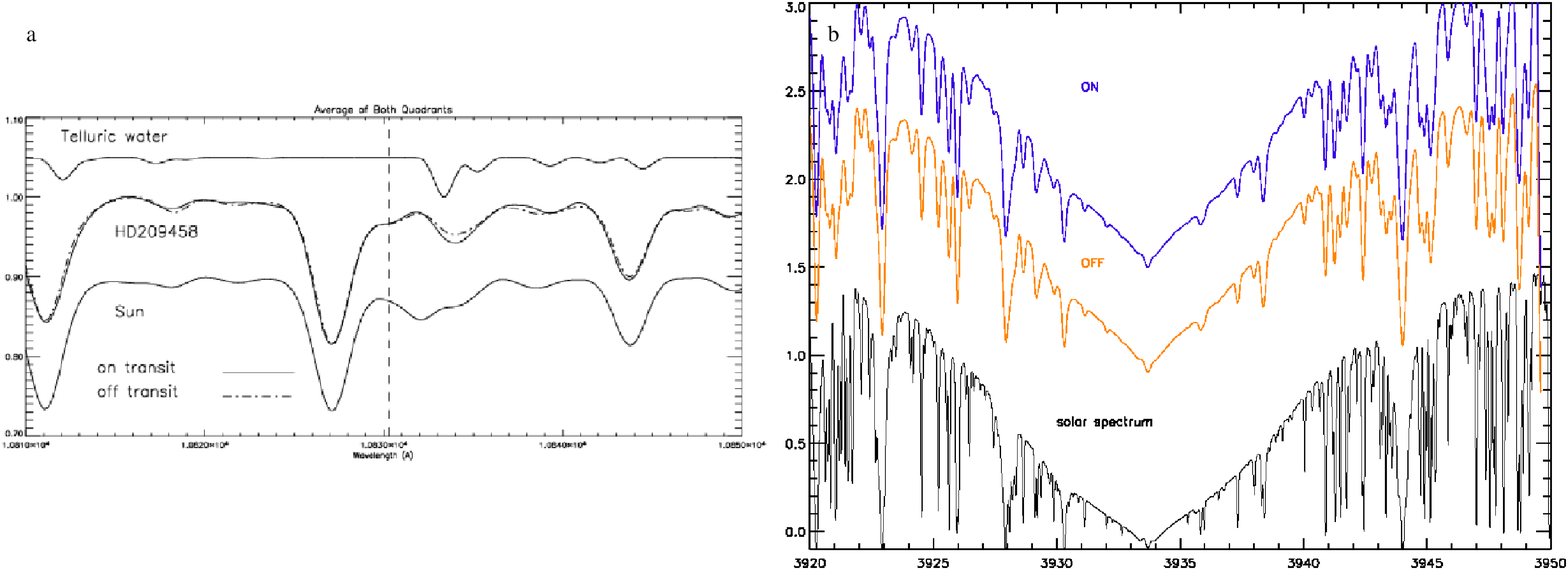}
\leavevmode
\caption{HD209458 Spectra. a: with ISAAC (Moutou et al. 2003), b: with UVES.}
\end{center}
\end{figure}

\section{Search for planet-induced chromospheric activity on HD209458}
Following the suggestion of Shkolnik et al.(2003), we searched signatures of a chromospheric activity induced by the
(magnetic more probably than tidal) heating of the star's outer atmosphere by the planet.
 
We have recovered high-resolution UVES spectra in the blue arm during one transit (and its off-transit) period in 2000.
The Ca II K at 3933\AA, Ca II K at 3968\AA\ and Al I at 3944\AA lines can be related to  chromospheric activity.
A sample of these spectra centered around the Ca II K feature is shown is Fig. 1b.
Our spectra show emission enhancement observed in the Ca II H band
but we found nothing significant in the two otherbands.
The variations could be indicative of the stellar chromospheric activity with the planet's magnetic field.

\section{Perspectives}

UVES data from the 5 transits in June-September 2002 are currently being analysed to look for exospheric planetary signature of other constituents.
Further and improved observations are required. We have time allocation this year.

\section{References}
Moutou, C., Coustenis, A., Schneider, J., Queloz, D. \& Mayor, M., 2003, in \aap, 405, 341 and references therein\\
Shkolnik, E., Walker, G.~A.~H, \& Bohlender, D.~A., 2003, submitted to \apj\\

\end{document}